\documentclass{aa}

\usepackage{graphicx}
\usepackage{natbib}
\usepackage{txfonts}

\begin{document}

\title {Galactic fountains and their connection with high and intermediate velocity clouds}

\author {E. Spitoni\inst{1}
\thanks {email to: spitoni@oats.inaf.it}
\and  S. Recchi\inst{2}
\and  F. Matteucci\inst{1, 2}}

\institute{ Dipartimento di Astronomia, Universit\`a di Trieste, via G.B. Tiepolo 11, I-34131  
\and  I.N.A.F. Osservatorio Astronomico di Trieste, via G.B. Tiepolo 11, I-34131}

\date{Received xxxx / Accepted xxxx}

\abstract{Sequential supernova explosions create supershells which can break out a stratified medium producing bipolar outflows. The gas of the supershells can fragment into clouds which eventually fall toward the disk producing the so called galactic fountains. }
{The aim of this paper is to calculate the expansion law and chemical enrichment of a supershell powered by the energetic feedback of a typical Galactic OB association at various galactocentric radii. We study then the orbits of the fragments created when the supershell breaks out and we compare their kinetic and chemical properties with the available observations of high - and intermediate - velocity clouds.}
{We use the Kompaneets (1960) approximation for the evolution of the superbubble driven by sequential supernova explosions and we compute the abundances of oxygen and iron residing in the thin cold supershell. We assume that supershells are fragmented by means of Rayleigh-Taylor instabilities and we follow the orbit of the clouds either ballistically or by means of a hybrid model considering viscous interaction between the clouds and the extra-planar gas.}
{Given the self-similarity of the Kompaneets solutions, clouds are always formed $\sim$448 pc above the plane. If the initial metallicity is solar, the pollution from dying stars of the OB association has a negligible effect on the chemical composition of the clouds. The maximum height reached by the clouds above the plane seldom exceeds 2 kpc and when averaging over different throwing angles, the landing coordinate differs from the throwing coordinate $\sim$ 1 kpc at most.}
{  The range of heights and $[O/Fe]$ ratios spun by our clouds suggest us that the high velocity clouds cannot have a Galactic origin, whereas intermediate velocity clouds have kinematic properties similar to our modeled clouds but overabundance observed for the $[O/Fe]$ ratios which can be reproduced only with initial metallicities which are too low compared for those of the Galaxy disk.}


\titlerunning{Galactic fountains}
\authorrunning{Spitoni et al.}
\maketitle

\section{Introduction}

Although significant progress has been made in the last few years in exploring the distribution and chemical composition  of intermediate and high-velocity clouds (IVCs and HVCs, respectively) in the halo of the Milky Way, an overall unified model for their formation is still lacking. These gas clouds are concentrations of neutral hydrogen (HI) with radial velocities which are not consistent with a simple model of differential galactic rotation (Richter et al. 2001). The distinction between HVCs and IVCs is loosely based on the observed radial velocities of the clouds; IVCs have radial velocities with respect to the Local Standard of Rest (LSR) of 30 km s$^{-1} \leqslant \vert  V_{LSR} \vert  \leqslant  $ 90 km s$^{-1}$ while HVCs have typical velocities $\vert  V_{LSR} \vert  >  $ 90 km s$^{-1}$. Several studies (Lu et al. 1998; Wakker et al. 1999; Murphy et al. 2000; Bluhm et al. 2001; Sembach et al. 2001) reveal different chemical compositions for several of these clouds in different directions in the sky. Oort (1970) proposed that these clouds represent condensed gaseous relics from the formation phase of the Milky Way. This idea was later revived by Blitz et al.\,(1999), who 
suggested that HVCs represent the building blocks of galaxies in a 
hierarchical galaxy formation scenario. 
Since the Galaxy is 
surrounded by smaller satellite galaxies (e.g., the 
Magellanic Clouds), another explanation is that IVCs and HVCs
are gaseous streams related to the
merging and accretion of these satellites by the Milky Way. 
In this picture, HVCs would be the gaseous counterparts of the Milky Way 
circumgalactic stellar streams, which are believed to 
represent the relics of dwarf galaxies that have been
accreted by the Milky Way (e.g., Ibata 1994). While all these models
assume that HVCs are truly extragalactic objects which are about to
merge with the Galaxy from outside, there are other scenarios 
that see the IVCs and HVCs as objects that have their 
origin in the disk of the Milky Way, e.g., as part of the 
so-called "galactic fountains". In the galactic fountain
model (Shapiro \& Field 1976; Houck \& Bregman 1990), hot gas is ejected
out of the Galactic disk by supernova (SN) explosions, and part of
this gas falls back in the form of condensed neutral
clouds which move at intermediate and high radial velocities.
Whatever the origin of the Milky Way IVCs and HVCs is, it
has become clear that they must play an important role in
the evolution of our Galaxy. Two extremely important parameters to distinguish between
the Galactic and extragalactic models of IVCs and HVCs
are the {\it distance} and the {\it chemical 
composition} of these clouds. In case of the Galactic model the clouds ejected by SN explosions must have a velocity high enough to explain the observed distances and their chemical composition must be correlated with the abundances of the gas in the disk 

 In this paper we examine the scenario in which the origin of these clouds is given by SN explosions, as the Galactic fountain model prescribes. We present precise calculations about the physical and chemical properties of the swept up gas by sequential SN events, in literature called supershell. Once the clouds leave the stellar disk, they move through the extra-planar gas halo. Two extreme types of models have been considered here for the motion of the gas through the extra-planar gas: i) the ballistic and  ii) the fluid homogeneous models. Numerous non-homogenous filamentary structures observed in the extra-planar halos (NGC891, NGC 5775) suggest that the kinematics of these halos could be interpreted with ballistic models. Ballistic models describe the gas as an inhomogeneous collection of clouds, subject only to the gravitational potential of the galaxy: for example, in the galactic fountain model the ejected gas from SN events falls back ballistically (Bregman 1980). These models are able to explain vertical motions
of the cold and warm gas components observed
in several spiral galaxies (e.g. Fraternali et al. 2004; Boomsma
et al. 2005). However, Collins et al. (2002) have tried a ballistic
model to describe the extra-planar ionized gas of NGC 891
and found problems in reproducing the observed kinematics.   
The ballistic model predicts that clouds migrate radially
outward as they cycle through the halo. The mass fluxes
estimated from the models of NGC 891 and NGC 5775
imply that significant amounts of gas can be involved in
these migrations. Such migrations could cause a redistribution
of gas that could affect metallicity gradients as well as
star formation properties. Such effects have been previously
investigated by Charlton \& Salpeter (1989), for example,
but extensive observations of the kinematic behavior of
edge-on galaxies should yield important constraints on such
redistribution. Bregman (1980) presented a model where the clouds ejected fall into the disk subject to the gravity of the Galaxy at the same coordinate where they were thrown. In this case the metallicity gradient of the Galaxy does not change.
 
Collins et al. (2002) suggested that the ballistic model problems could be solved by
considering the presence of drag between disk and halo, such
as through magnetic tension or viscous interactions between clouds. Barnab\`e et al. (2006) presented fluid stationary models  able  to reproduce the observed negative vertical gradient of the rotation velocity of the extra-planar gas. In these models the gas is described as a stationary rotational fluid. In the case of a single cloud, the solutions of this model predict the drag invoked by Collins et al. (2002), as ram pressure between cloud and fluid halo. This suggests that a correct description of the extra-planar gas dynamics may be found in hybrid ballistic-fluid stationary models. Recently, Fraternali \& Binney (2008) presented a new formulation for viscous cloud-halo interactions in the framework of the cold gas accretion onto spiral galaxies in the local universe. 

In this paper we will follow the clouds ejected by SN explosions both in the Galaxy in the purely ballistic model and in the hybrid one. The paper is organized as follows: in Sect. \ref{model} we will describe our model, in Sect \ref{results} we will show our results concerning the chemical abundances of the supershells and the cloud orbits in the purely ballistic and in the hybrid model. In Sect. \ref{hvc} we compare our results with observations of the HVC Complex C and the IVC VIArc. In Sect. \ref{conclusions} we report our conclusions.

\section{Description of the model}\label{model}

Type II SNe usually occur in OB associations containing several dozens of massive stars. Sequential SN explosions create a superbubble, whereas the swept up gas is concentrated in a thin cold shell called supershell. The superbubble expansion in a stratified medium does not follow a spherical evolution and the Kompaneets (1960) approximation well describes the way in which the superbubble grows up in the meridional  Galaxy plane. After some interval of time, a region of the supershell can fragment due to the occurrence of the Rayleigh-Taylor instabilities (RTIs), therefore clouds of gas can form. Once left the stellar disk, the orbit of each cloud can be followed either ballistically or with a hybrid model considering viscous interaction between the cloud and the extra-planar gas.

\subsection {The Kompaneets approximation}
\begin{figure}[!h]
\begin{center}
\includegraphics[width=0.30\textwidth]{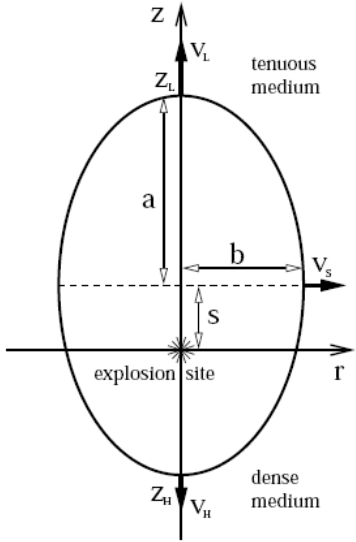} 
\caption{Explanatory picture of the superbubble shape (see Kompaneets 1960) under the assumption that the density  below the plane is constant (see the Text). There are reported some physical quantities: the height of the superbubble in the exponential atmosphere $(Z_L)$ and below tha plane $(Z_H) $, and the semi-minor axis $(b)$. Figure taken from Kompaneets (1960).}
 
\label{bubble}
\end{center}
\end{figure}

\begin{figure}[!h]
\begin{center}
\includegraphics[width=0.45\textwidth]{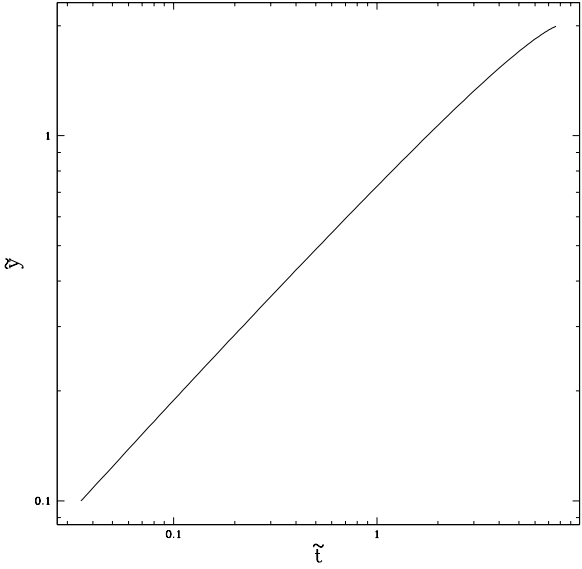} 
\caption{Evolution in the dimensionless space of the quantity $\tilde{y}$ (eq. \ref{y}) in function of $\tilde{t}$. $\tilde{y}$ is related to the top of the superbubble ($Z_{L}$) evolution from the eq. (\ref{zlh}). The dimensional time $t$ is given by:\ $t=\tilde {t} \times (\rho_{0}H^{5}/L_{0})^{1/3}$}  
\label{fy}
\end{center}
\end{figure}

\begin{figure}[!h]
\begin{center}
\includegraphics[width=0.45\textwidth]{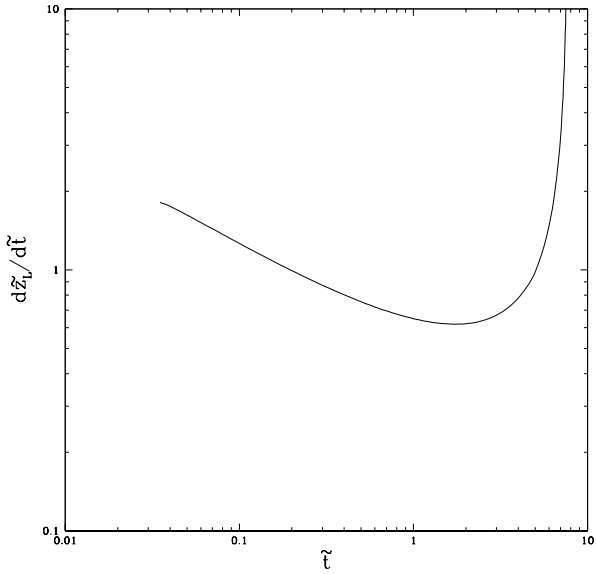} 
\caption{Evolution in the dimensionless space of the velocity  $d\tilde{z_{L}}/d\tilde{t}$ as function of $\tilde{t}$. We note that the shell starts accelerating at $\tilde {t_{a}}=1.72$ .}  
\label{fdz}
\end{center}
\end{figure}

 The physical parameters of the superbubble can be fairly well described by the Kompaneets (1960) model. The Kompaneets approximation assumes the following : uniform pressure within the superbubble, superbubble expansion in a direction normal to the local surface and expansion speed implied by a strong shock (i.e., the internal pressure dominates the external
pressure). Kompaneets found an analytic expression for the shape of the bubble during its expansion in an exponential atmosphere with density:

\begin{equation}
\rho(z)=\rho_{0}exp(-z/H),
\label{rhozp}
\end{equation}
where $\rho_{0}$ and $H$ are the disk density and height scale, respectively.
We consider the shock expansion in cylindrical coordinates $(z,R)$. In the work of Kompaneets it was assumed that for $z<0$ the ISM was denser than the gas residing above the disk. Therefore, in this approximation, the bubble has an asymmetric evolution with respect to the plane of the galaxy. Kompaneets found analytic expressions for the top and bottom sites of the remnant. Referring to Fig. \ref{bubble} we have:
\begin{equation}
z_{L,H}=-2H\ln \left( 1 \mp \frac{y}{2H}\right),
\label{zlh}
\end{equation}
and
\begin{equation}
b=2H\arcsin \left( \frac{y}{2H} \right).
\end{equation}
The quantity $y$ is a transformed variable (with units of length) defined by:
\begin{equation}
y= \int_{0}^{t}\sqrt{\frac{\gamma+1}{2}\frac{E_{th}}{\rho_{0}\Omega}}dt,
\label{y}
\end{equation}
where $\Omega$ and $E_{th}$ and $\gamma$ are the volume, the thermal energy  and the ratio of specific heats for the superbubble, respectively . The thermal energy is calculated from the differential equation:
\begin{equation}
\frac{dE_{th}}{dt}=L_{0}-P\frac{d\Omega}{dt},
\label{E}
\end{equation}
where $L_0$ is the wind luminosity and the pressure is given by the equation of state:
\begin{equation}
P=(\gamma -1)E_{th}/\Omega.
\label{P}
\end{equation}
In our work for the ISM z-profile we assume an exponential law also for $z<0$. Hence the superbubble evolves symmetrically with respect to the galactic plane resulting in a peanut-like shape, as suggested by Mac Low et al. (1988) and by Tenorio-Tagle et al. (1999). The volume of the superbubble can be estimated by:
\begin{equation}
\Omega=2\cdot\frac{4}{3}\pi a^{2}b- \frac{4}{3}\pi|z_{H}|^{3}
\label{Omega}
\end{equation}
where $a=(z_{L}+|z_{H}|)/2$ is the semi-major axis.
A numerical integration of eqs. (\ref{y}) and (\ref{E}), with the help of eqs. (\ref{P}) and (\ref{Omega}),     yields $y(t)$ and $E_{th}$ which implicitly give the time evolution of the superbubble. The solution depends on only three parameters: the scale height $H$, the density near the source $\rho_{0}$, and the wind luminosity $L_0$. Hence the problem is solved most naturally in a system of units which are determined by these three parameters. The unit of length is then $H_{0}$, the unit of mass is $\rho_{0}H^3$, and the unit of time is $(\rho_{0}H^{5}/L_{0})^{1/3}$. We obtain the dimensionless solution by integrating the dimensionless form of eqs. (\ref{y}) and (\ref{E}). In Figs. \ref{fy} and \ref{fdz} we report the dimensionless evolution of $y$ and the velocity $dz_{L}/dt$.  The dimensionless analysis predicts that the superbubble begins to accelerate at (see Fig. \ref{fdz}):
\begin{equation}
\tilde{t_{a}}=\frac{t_{a}}{(\rho_{0} H^{5}/L_{0})^{1/3}}=1.72,
\label{tilde}
\end{equation}
namely when $dz_{L}/dt$ has a minimum. 
 We underline that the simulated superbubbles, characterized by different values of $\rho_0$, $H$, $L_0$,  reach the accelerated phase at different times, as shown in eq. (\ref{tilde}), but they have the same shape and volume. This is a direct consequence of the self-similar nature of the Kompaneets solutions. Using the scale height $H$=141 pc (see Sect. \ref{galaxy}) we obtain for the height of the supeshell and for the minor semiaxis the values $z_{L}=191pc$ and $b=145pc$. From the Fig. \ref{fdz} we note that, after the beginning of the accelerated phase, the velocity remains roughly constant for $\sim$ 2 $\tilde{t}$. In this interval of time the acceleration is therefore negligible. Therefore we assume that the RTIs begin to grow when the velocity is increased by a factor of 10\%, because from that moment on we can assume that the acceleration is not negligible. The choice of this percentage is arbitrary but in this way our results are consistent with the work of Mac Low \& McCray (1988). We have also tested velocity variations of 5\% and 20\% and the results are not very different to the ones presented in this paper. In our model we assume that owing to the RTIs, the supershell fragments and we consider each fragment as a cloud with initial velocity given by the supershell velocity at the moment of the fragmentation. RTIs arise when the supershell accelerates since at that moment, the effective gravity of the supershell exceeds the one of the tenuous superbubble.  From Tenorio-Tagle et al. (1987) we have taken an estimate of the time of growth of these highly irregular structures:
\begin{equation}
t_{R-T}=10^5 R_{40}^2 \sqrt{\frac{\lambda_{10}}{E_{51}(t_{a}+t_{\Delta v})}n_{0}         }\mbox{ } \mbox{ yr},
\end{equation}
where $\lambda_{10}$ and $R_{40}$ are perturbation wavelength and remnant radius measured in units of 10 and 40 pc, respectively, $E_{51}$ is the energy measured in units of $10^{51}$ erg and $t_{\Delta v}$ is the time necessary for reaching an appreciable acceleration. At the time when RTIs start growing, the supershell has reached $z_{L}$=312 and $b$=228. We assume for simplicity the formation of a single cloud that corresponds to the maximum perturbation wavelength for the thin shell, roughly given by the supershell thickness. In the case of adiabatic expansion of a spherical thin shell, the thickness $\delta$ is given by $\delta =R/12$, where $R$ is the radius of the shell. We assume for simplicity that $\delta$ is the maximum perturbation wavelength and we make the approximation that $R=z_L$, therefore, $\delta=\lambda$ = 26 pc. If we define $\tilde{t}_{R-T}$ as:
\begin{equation}
\tilde{t}_{R-T}=\frac{t_{R-T}}{(\rho_{0} H^{5}/L_{0})^{1/3}}
\label{}
\end{equation}
we obtain the total adimensional time  necessary for the growing of instabilities and for the fragmentation of the bubble:
\begin{equation}
\tilde{t}_{final}=\tilde{t_{a}}+\tilde{t}_{\Delta v}+ \tilde{t}_{R-T}=4.37,
\label{}
\end{equation}
Considering different values of the parameters $\rho_0$, $H$, $L_0$, we obtain different times at which the cloud could leave the disk:
\begin{equation}
t_{final}=4.37(\rho_{0} H^{5}/L_{0})^{1/3}.
\label{final}
\end{equation}
Since the internal pressure dominates the external pressure, the expansion speed is that given by the Hugoniot condition for a strong shock: 
\begin{equation}
v_{n}=\sqrt{\frac{\gamma+1}{2}\frac{P(t)}{\rho(z)}}.
\end{equation}
  
\subsection{Abundances of Fe and O in the superbubble}

In this section we compute the total amounts of $O$ and $Fe$ that reside in the supershell at the time $t_{final}$ (see eq.\ref{final}) in which we suppose the fragmented shell can leave the stellar disk of the Galaxy. We assume that all the stars with masses larger than 8 $M_{\odot}$ explode as Type II SNe and we adopt the following main sequence lifetimes (Matteucci \& Greggio 1986):
 \begin{equation}
 \tau_{m} =1.2m^{-1.85}+0.003 \mbox{Gyr}.
\label {taum}
\end{equation}
 In our model, the upper mass limit of the OB association is assumed to be 40$M_{\odot}$. The lifetime for a 40$M_{\odot}$ is, from eq. (\ref{taum}),  $\tau_{40} \simeq 4.30 \times 10^6$ yr. 
  
For any parameter choice, at the time in which the shell fragments and leaves the disk, the smaller mass stars in the OB association are still alive. Using eq. (\ref{taum}) we are able to estimate the lower mass limit for a SN event corresponding to the time at which the cloud forms. The time at which the cloud is formed is given by $t_{final}$ plus the lifetime of the largest considered mass, namely:
\begin{equation}
 \tau_{lim}=t_{final} +\tau_{40}.
\end{equation}
 The mass of metals (in our model we consider only $O$ and $Fe$) ejected by the range of SNe considered is given by:

\begin{equation}
\label{MZ}
 M_{el_{\star}}=\int_{M_{inf}}^{40}m_{el}(m) \phi(m) dm,
\end{equation}
where $\phi(m)$ is the Initial Mass function (IMF) assumed to be the Salpeter (1955) one, namely: $\phi (m)=A m^{-(2.35)}$. A is the normalization constant, given by:
 \begin{equation}
 \int_{8}^{40} \phi (m)dm=A \int_{8}^{40}m^{-2.35}dm= SNe,
\end{equation}
where $SNe$ is the assumed number of SNe in the OB association.
In eq. (\ref{MZ}) the term $m_{el}(m)$ is the ejected mass of the $el$ element considered ($O$ or $Fe$) by the star of initial mass $m$ (Woosley \& Weaver 1995). We consider the stellar yields at four different metallicities: solar metallicity $Z_{\odot}$, 0.1 $Z_{\odot}$, 0.01 $Z_{\odot}$, $10^{-4}Z_{\odot}$. We assume that the ISM in the disk has the same  metallicity of the OB association. For the solar abundances (by mass) we use the Anders \& Grevesse (1989) values:

\begin{equation}
\label{MF}
O_{16_{\odot}}=9.59 \times 10^{-3} \mbox{ ,  } \mbox{   } Fe_{56_{\odot}}=1.17 \times 10^{-3}.
\end{equation}
We use this set of solar abundances in order to be consistent with the yields of Woosley \& Weaver (1995) which refer to Anders \& Grevesse (1989) abundances. However it should be recalled that recently Asplund et al. (2005) recomputed the $O$ solar abundance and deduced a $\log (O/Fe)_{\odot}$ lower by $\sim$ 0.24 dex than the Anders \& Grevesse (1989) one. The total amount $M_{shell-el}$ of the mass of the element which resides in the thin shell of the superbubble is given by the eq.:

\begin{equation}
\label {masshell}
 M_{shell-el}=M_{shell}\cdot Z +  M_{el_{\star}},
\end{equation}
where $Z$ is the initial metallicity of the OB association (and consequently of the disk ISM).
We adopt the extreme approximation that total amount of the ejected metals from SNe ends up in the shell. $  M_{shell}$ is the mass of the ISM swept up into the thin shell and is given by:

\begin{equation}
\label {MzP}
  M_{shell}(z>0)=\pi \rho_{0} b^2 \int_{0}^{z_{L}}e^{-z/H}\left( 1-\left(\frac{z-a+|z_{H}|}{a}\right)^2\right) dz.
\end{equation}

We underline that for the $Fe$ we use only solar yields for all range of metallicities (Chiappini et al. 2001).

\subsection{ Galaxy model}
\label{galaxy}
The potential well of the Galaxy is assumed to be the sum of three components: a dark matter halo, a bulge and a disk.
The dark matter halo gravitational potential is assumed
to follow the Navarro, Frenk and White (1996) profile:
\begin{equation}
\Phi_{dm}(r)=-4 \pi r_{dm,0}^{2} \, \rho_{dm,0} \frac{\ln (1+x)}{x}, 
\end{equation}
where $ \rho_{dm,0}$ is a reference density, $r_{dm,0}$ is a scale radius,
$x = r/r_{dm,0}$ and $r$ is the spherical radius. The halo is truncated
at a radius  $r_{dm,t}$ beyond which its potential follows
the 1/r profile. 
The bulge gravitational potential is given by (Hernquist 1990):

\begin{equation}
\Phi_{b}(r)=-\frac{GM_{b}}{r_{b,0}+r} 
\end{equation}
where $r_{b,0}$ is a scale radius and $M_{b}$ is the bulge mass. For the disk potential we have chosen the axisymmetrical Miyamoto \& Nagai (1975) model, which in cylindrical coordinates $(R,z)$ can be written as:
\begin{equation}\label{eq:fi}
\Phi_M(R,z)=-\frac{GM_d}{\sqrt{R^2+{\left(a+\sqrt{z^2+b^2}\right)}^2}}.
\end{equation}
Tab. (\ref{GALAXYp}) gives the values of all the parameters concerning the Galaxy model. For the Interstellar Medium (ISM) z-density profile we used the eq. (\ref{rhozp}) where: $\rho_{0}=n_{0}\mu m_{p}$ is the density in the disk plane; $m_{p}$ is the proton mass and $\mu $ is mean molecular weight for the disk (assumed to be 0.61). For the value of $H$ we adopted the vertical  distributions of the various interstellar components (molecular, cold HI, warm HIa, warm HIb, HII Regions, and diffuse HII) in the solar neighborhood as reported by Cox (2005). We obtained:
\begin{equation}
H=\frac{1}{\rho_{0}} \int_{0}^{\infty}\sum_{i=1}^{6}\rho_{i}(z)dz\simeq 141 \mbox{pc}.
\end{equation}
The ISM density profile along the radius $R$ of the Galaxy is taken by Wolfire et al. (2003). In our model we consider the presence of an extra-planar gas halo. This halo can be described by means of the perfect gas law:

\begin{equation}\label{eq:p}
P=\frac{{\rho}{kT}}{{\mu}{m_p}}. 
\end{equation}
For simplicity we assume a isothermal and self-gravitant distribution, hence introducing the parameter:

\begin{equation}\label{eq:bet}
{\beta}_0=\frac{kT_0}{{\mu}{m_p}},  
\end{equation}
 by means of which the eq. (\ref{eq:p}) can be written as:
\begin{equation}\label{eq:po}
P={\beta}_o\rho.
\end{equation}
Using the hydrostatic equilibrium equation with the assumption of a static halo of gas we have:

\begin{equation}\label{eq:rog}
\rho(R,z)={\rho}_o e^{(-\Phi_{tot}+\Phi_0)/\beta_0},
\end{equation}
where $\Phi_{tot}= \Phi_{dm}+\Phi_{b}+\Phi_M$ and $\Phi_0$ is the value of $\Phi_{tot}$ calculated in the galactic center. The temperature $T_o$ has been chosen following the relation:
\begin{equation}
\frac{3}{2}kT_o=\frac{3m_p\mu v_{c8}^2}{2},
\end{equation}
where $v_{c8}$ is the circular velocity in the plane of Galaxy in the solar neighborhood and we obtain that:
\begin{equation}
T_0=3.7\times10^{6\mbox{}}\mbox{K}
\end{equation}
 for the chosen parameters.

\begin{table*}[htp]
\caption{Galactic parameters. }
\label{GALAXYp}
\begin{center}
\begin{tabular}{cccccccc}
  \hline
\noalign{\smallskip}



 $\rho_{dm,0}$ &$M_{b} $&$M_d $  &$r_{dm,0} $& $r_{dm,t} $&$r_{b,0}$   &$a$&$b$\\
  $10^{-24}$ g cm $^{-3}$ &$10^{10} M_{\odot}$ &$10^{10} M_{\odot}$ &kpc&kpc &kpc &kpc &kpc\\
\noalign{\smallskip}

\hline
\noalign{\smallskip}

0.29&3.5 &7.69 &30.8&347.7 &0.8&8.45&0.26\\
\noalign{\smallskip}

\hline
\end{tabular}
\end{center}
\end{table*}

\subsection{ Galactic fountains in a ballistic model}

Once the top of the supershell reaches the height above the galactic plane related to the time $t_{final}$ (see eq. \ref{final}), the thin shell can leave the stellar disk and move towards the extra-planar gas. Ballistic models describe the gas as an inhomogeneous collection of clouds, subject only to the gravitational potential of the Galaxy.
 The fragments of the shell  have different initial  velocity modulus   $\parallel \mathbf{v}_{o}\parallel=v_{n}$ and masses depending on the the number of SNe in the OB association and the initial throwing radial coordinate. We have chosen for our simulation Cartesian coordinates with versor $( \mathbf{\hat{e}}_x,\mathbf{\hat{e}}_y,\mathbf{\hat{e}}_z    )$.
 Given a generic point in this space $(x,y,z)$, it corresponds in the  meridional plane of the Galaxy to $(R,z)=(\sqrt{x^2 +y^2},z)$, where $R$ is the radial coordinate. Since the $\parallel \mathbf{v}_{o}\parallel$ velocities are relative to the local standard of rest for the trowing radial  coordinate $R_{0}$, we have that in the inertial reference frame of the simulation in Cartesian coordinates:

\begin{equation}
\mathbf{v}_{initial}(R_{0})= \mathbf{v}_{o}+v_{c}(R_{o},0)\mathbf{\hat{e}}_{y}.
\end{equation}                   
For each choice of $\parallel \mathbf{v}_{o}\parallel$ we consider several throwing directions  to take into account the possibility that the
breakup of a single supershell produces several fragments, each one with its own orbit. Given a throwing direction, a generic vector $\mathbf{v}_0$ has components along $\mathbf{\hat{e}}_x$, $\mathbf{\hat{e}}_z$ e $\mathbf{\hat{e}}_y$ with  respect to our inertial frame:   
\begin{equation}\label{eq:getto}
\left\{
\begin{array}{ccc}
v_{z_{init}}=v_n{\cos}{\gamma}&&\\
\\
v_{x_{init}}=v_n{\sin}{\gamma}{\cos}{\beta}&&\\
\\
v_{y_{init}}=v_n{\sin}{\gamma}{\sin}{\beta}+v_c(R_0,0)&&\mbox{.}
\end{array}
\right.
\end{equation}
The parameters $\gamma$ e $\beta$ vary in order to recreate, in the local standard of rest, a  symmetrical fountain of jets:
\begin{equation}
\beta_{i}=\frac{i\pi}{4}\mbox{ with }i=0,1,..7
\end{equation}
and
\begin{equation}
\gamma_{j}=\frac{\pi}{2}-\frac{\pi}{1+j} \mbox{ with }j=2,3,..5.
\end{equation}
We consider also the case in which  $\mathbf{v}_0$ has only a component along $\mathbf{\hat{e}}_{z}$. In conclusion, for each velocity modulus $v_n$ we calculate a fountain composed by 33 jets.

\subsection{Beyond the ballistic model: hybrid ballistic-fluid stationary model}

In this paper we want to provide some hydrodynamical modifications to previous ballistic model. In the motion equations for a single cloud we insert a Stokes term in order to consider also the viscous interactions between cloud and extra-planar gas. We introduce the drag time $ t_{drag}$, i.e the time required for an HI cloud to corotate with a homogeneous gas halo (Barnab\`e et al. 2005):
\begin{equation}
t_{drag}=\frac{8}{3C_D}\frac{r_c}{v_{rel}}\chi,
\end{equation}
where $C_D\simeq1$ is a numerical coefficient, $r_c$ is the radius of a typical  H {\small I} cloud, $v_{rel}$ is the modulus of the relative velocity between the cloud and the homogeneous extra-planar gas, and $\chi\equiv \rho_c/\rho$ is the ratio between the cloud and the medium densities.
 $t_{drag}$ was estimated  assuming pressure equilibrium between cold and hot components, from which $\chi=T/T_c\approx 3000$. A fiducial value of the relative velocity is obtained by assuming  $v_{rel}\approx2\times10^7$ $\mbox{cm}\mbox{ }{s}^{-1}$, while the cloud radius $r_c$ is estimated as:
\begin{equation} 
4\pi{r_c}^3\rho_c/3=10^5M_\odot\tilde{M_5},
\end{equation}
where  $\tilde{M_5}$ is the cloud mass in units of  $10^5$ $M_\odot$, thus following the work of Barnab\`e et al. (2005) one obtains: 



\begin{equation}
t_{\rm drag} \approx      2.7 \times 10^8\left( 
                {\tilde{M}_5\over n_{\rm p}} 
                \right)^{1/3}\textrm{yr},
\label{tdrag2}
\end{equation}
where $n_p$ is the numerical density of the extra-planar gas with the density profile described by eq. (\ref{eq:rog}). 
We note that a viscosity term could not be directly inserted in a Lagrangian formulation, thus we must write the explicit {\itshape drag} term to add in the motion equations. The {\itshape drag} term  can be derived from the brake force estimate as:

\begin{equation}
\mathbf{a}_{drag}\equiv-\frac{\mathbf{v}-\mathbf{v}_{{g}}}{t_{drag}},
\end{equation} 
where $\mathbf{v}$ is the velocity of the cloud whereas $\mathbf{v}_{{g}}$ is the velocity for the extra-planar gas that we assumed equal to zero.

\section{Our results}\label{results}
\subsection{Chemical composition of the clouds}\label{clouds}
In our models we vary the number of SNeII in the OB association ($SNe$) and the throwing radial coordinate ($R_0$). We consider four possible OB associations containing 10, 50, 100, 500 SNe respectively. Assuming an explosion energy of $10^{51}$ erg, the luminosities $L_0$ of these OB associations are $10^{37}$, $5 \times 10^{37}$, $10^{38}$ and $5 \times 10^{38}$ ergs$^{-1}$ respectively. These numbers of massive star in OB associations are consistent with the observations (de Zeeuw et al. 1999). We simulate galactic fountains with 3 different throwing radial coordinates $R_{0}$: 4, 8, 12 kpc. Given the assumed Galaxy model, $R_0$ defines the disk density $\rho_0$, whereas the scale height is constant (see Sect \ref{galaxy}).
 In Tabs. \ref{ta4}, \ref{ta8}, \ref{ta12} we summarize the results for fragmentation times and the velocities of the superbubbles in the direction perpendicular to the galactic plane at those times for 4 kpc, 8 kpc,12 kpc, respectively.

\begin{table}[htp] 
\caption{Cloud formation times and cloud velocities for $R_0=4$ kpc.}
\label{ta4}
\begin{center}
\begin{tabular}{ccc}
  \hline\hline
\noalign{\smallskip}

 SNe& $t_{final}$   & $v_{n}$ \\

 & [yr]  & [kms$^{-1}$] \\
\noalign{\smallskip}
  \hline 
  10 & 2.29$\times 10^{7}$  & 23\\
 50 & 1.34$\times 10^{7}$  &  39\\
 100 & 1.06$\times 10^{7}$ &  49 \\
 500 & 6.21$\times 10^{6}$ &  83 \\

  \hline
 \end{tabular}
\end{center}
\end{table}

\begin{table}[htp]
\caption{Same of Tab. \ref{ta4} but for $R_0=8$ kpc.}
\label{ta8}
\begin{center}
\begin{tabular}{ccc}
  \hline\hline
\noalign{\smallskip}

 SNe& $t_{final}$   & $v_{n}$ \\

 & [yr]  & [kms$^{-1}$] \\
\noalign{\smallskip}
  \hline 
  10 & 1.90$\times 10^{7}$  & 27\\
 50 & 1.11$\times 10^{7}$  &  46\\
 100 & 8.84$\times 10^{6}$ &  58 \\
 500 & 5.17$\times 10^{6}$ &  100 \\

  \hline
 \end{tabular}
\end{center}
\end{table}

\begin{table}[htp]
\caption{Same of Tab. \ref{ta4} but for $R_0=12$ kpc.}
\label{ta12}
\begin{center}
\begin{tabular}{ccc}
  \hline\hline
\noalign{\smallskip}

 SNe& $t_{final}$   & $v_{n}$ \\

 & [yr]  & [kms$^{-1}$] \\
\noalign{\smallskip}
  \hline 
  10 & 1.67$\times 10^{7}$  & 31\\
 50 & 9.77$\times 10^{6}$  &  53\\
 100 & 7.74$\times 10^{6}$ &  67 \\
 500 & 4.53$\times 10^{6}$ &  114 \\

  \hline
 \end{tabular}
\end{center}
\end{table}
For all the values of $SNe$ and $R_0$ considered at the time $t_{final}$, at which clouds are thrown out of the disk, the supershell presents: $z_{L}=448$ pc, $z_{H}=165$ pc and $b=259$ pc. Our results are in agreement with the work of Mac Low \& McCray (1988): instabilities, for roughly the same luminosity range, become important at $3H$ height scale.  The total mass of gas swept up by the SN shock wave for positive z-coordinates is given by the eq. (\ref{MzP}).

\begin{figure}[ht]
\begin{center}
\includegraphics[width=0.45\textwidth]{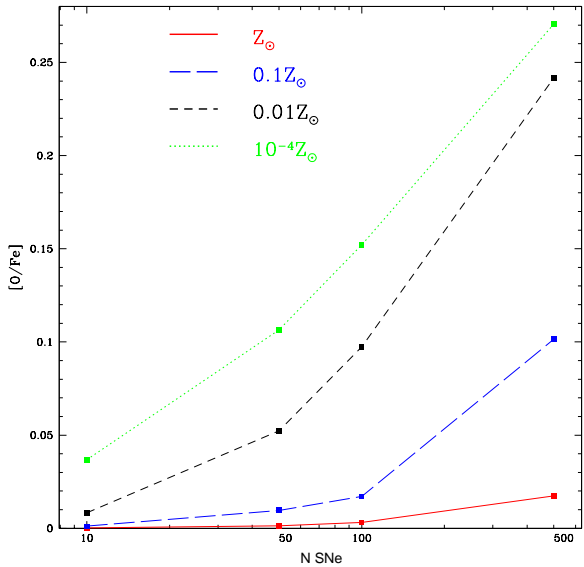} 
\caption{[O/Fe] ratios as functions of the number of SNe and  of different disk gas metallicities in the case of $R_0$=8 kpc using stellar yields given by Woosley \& Weaver (1995).}  
\label{result}
\end{center}
\end{figure}

\begin{figure}[ht]
\begin{center}
\includegraphics[width=0.45\textwidth]{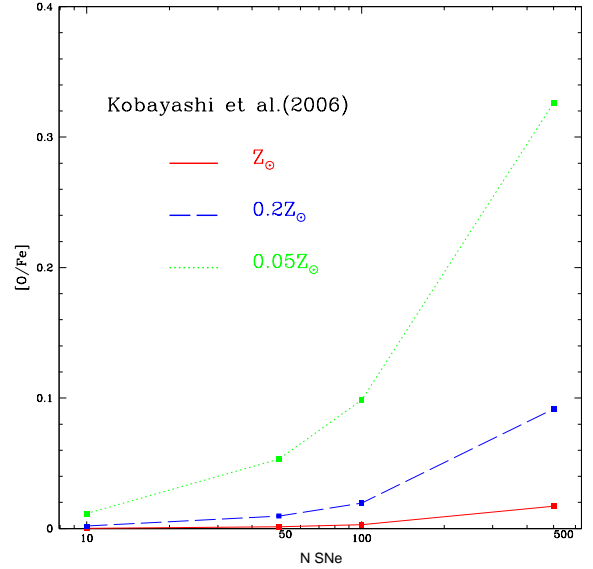}
\caption{As in Fig.\ref{result} using stellar yields given by Kobayashi et al. (2006).}  
\label{resultk}
\end{center}
\end{figure}

\begin{figure}[ht]
\begin{center}
\includegraphics[width=0.45\textwidth]{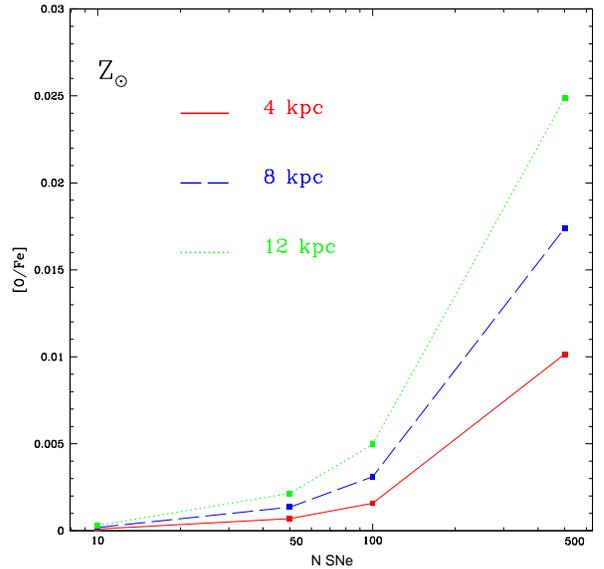}
\caption{[O/Fe] ratios as functions of the number of SNe and the throwing radial coordinate in the case of solar metallicity using stellar yields given by Woosley \& Weaver (1995).}  
\label{zsol}
\end{center}
\end{figure}

\begin{figure}[ht]
\begin{center}
\includegraphics[width=0.45\textwidth]{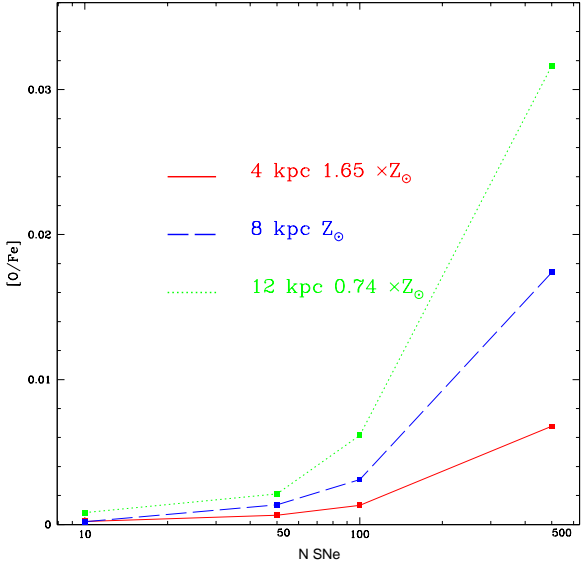}
\caption{[O/Fe] ratios as functions of the number of SNe and the initial throwing radial coordinate, taking for the ISM metallicities the average observed values along the Galactic disk from Cepheids by Andreievsky et al. (2002a-c, 2004) and Luck et al. (2003) (see Cescutti et al. 2006).}  
\label{zreal}
\end{center}
\end{figure}

In conclusion, we obtain that the masses of the ISM swept up into the thin shell for $R_0$= 4, 8, 12 kpc, respectively are:

\begin{equation}
\label {}
 M_{4}(z>0)=10.07 \times 10^5  M_{\odot},
\end{equation}

\begin{equation}
\label {}
 M_{8}(z>0)=5.79 \times 10^5  M_{\odot},
\end{equation}

\begin{equation}
\label {M12}
 M_{12}(z>0)=3.89 \times 10^5  M_{\odot}.
\end{equation}
 All the results about the $O$ and $Fe$ abundances in the clouds ejected by sequential SN explosions as functions of $SNe$ and $R_0$ are reported in the Online material. $M_{*Fe_{56}} $ and $M_{*O_{16}} $ are the total amounts of $Fe_{56}$ and $ O_{16}$ in unit of $M_{\odot}$, whereas $ M_{\star ej}$ is the total mass ejected by the OB association:
\begin{equation}
\label{Mtot}
 M_{\star ej}=\int_{M_{inf}}^{40}m_{tot}(m) \phi(m) dm,
\end{equation}
where $m_{tot}(m)$ is the total mass ejected by a SN as a function of its initial mass and metallicity.  $M_{tot}$, $ X_{*Fe_{56}}$,  $X_{*O_{16}}$ and $[O/Fe]$ are:
\begin{equation}
\label{Mtot2}
 M_{tot}=M_{shell}+ M_{\star ej},
\end{equation}
\begin{equation}
\label{X}
 X_{*Fe_{56}}= \frac{M_{Shell  Fe_{56}}}{M_{tot}}     \mbox{  ,  } X_{*O_{16}}= \frac{M_{Shell  O_{16}}}{M_{tot}}.
\end{equation}

\begin{equation}
\label{O/fe}
 [O/Fe]=\log \left( \frac{M_{Shell O_{16}}}{M_{Shell Fe_{56}}} \right)- \log \left( \frac{O_{16_{\odot}}}{ Fe_{56_{\odot}}} \right)  .
\end{equation}
In Fig \ref{result} we show the predicted $[O/Fe]$ ratio as a function of the number of SNe and of the initial metallicity in the solar vicinity. In the meridional plane of the Galaxy the initial conditions are $(R,z)$=( 8 kpc, 448 pc). We note that significant over-abundances of $O$ relative to $Fe$ are found only in the case of a large number of SNe and low initial metallicity. In Fig. \ref{resultk} we report the same quantities but using stellar yields given by Kobayashi et al. (2006). In Fig. \ref{zsol} we show $[O/Fe]$ ratios as functions of the number of SNe assuming solar metallicity but varying the initial throwing coordinate $R_o$. We note that larger radial coordinates yield a larger $[O/Fe]$, because the amount of the swept-up pristine gas is smaller (see eq. \ref{M12}) and therefore the new $\alpha$ elements ejected by SNe are less diluted.  In Fig. \ref{zreal} we report the $[O/Fe]$ ratios varying the initial throwing coordinate $R_0$ and taking for the initial ISM metallicities the average observed values given by Andreievsky et al. (2002a-c, 2004) and Luck et al. (2003), as a function of galactocentric distance, by analyzing Galactic Cepheids (see Cescutti at al. 2006). Referring to the Tab. 4 of the work of Cescutti et al. (2006), we find: $Z= 1.65 \times Z_{\odot}$ for $R_0=4$ kpc and $Z= 0.74 \times Z_{\odot}$ for $R_0=12$ kpc.

\subsection{Dynamics of the galactic fountains}\label{fountains}

\begin{figure}
\includegraphics[width=0.45\textwidth]{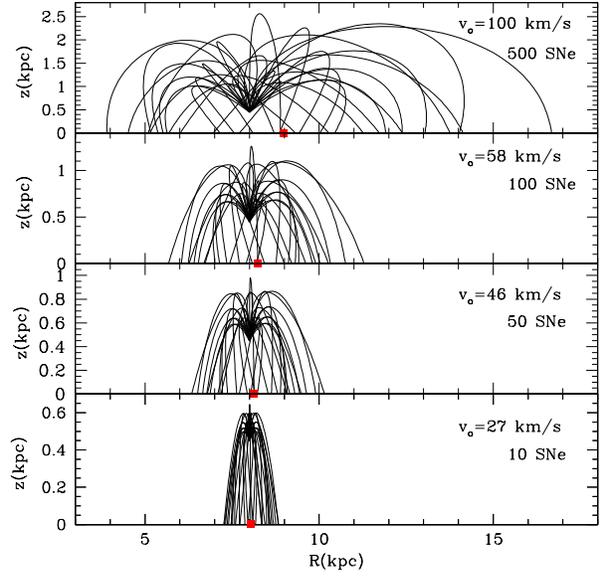}
\caption{Galactic fountains reported in the meridional plane in the purely ballistic model with the same spatial initial conditions: $(R,z)$=(8 kpc, 448 pc). Squares on the $R$ axis are the average falling radial coordinate. }
\label{8kpcs}
\end{figure}

Our analysis focuses first on the study of solar neighborhood galactic fountains (i.e. with $R_0=8$ kpc). In Fig. \ref{8kpcs} we show our results concerning the orbits of the shell fragments once they leave the stellar disk in the purely ballistic model. The trajectories of the orbits of each galactic fountain are plotted in the meridional plane of the Galaxy $(R,z)$. Spatial initial conditions are the same for all cases considered: $R_{0}=8$ kpc and $z_{0}=448$ pc, whereas the initial velocities depend on the number of SNe considered (Tab. \ref{ta8}). In  Fig. \ref{8kpcs}, squares on the $R$ axis reported the average falling radial coordinate in the Galaxy plane. As we can see, the clouds are preferentially thrown outwards, but their final average landing coordinates differ by 1 kpc at most from the throwing coordinate.  This result is consistent with the works of Bregman (1980), Fraternali \& Binney (2008) and Melioli et al. (2008). On the other hand, in their recent hydro-simulations for the Milky Way disk Booth \& Theuns (2007) found that clouds ejected from galactic fountains return to the disk  at average galactocentric 
distances  several times  larger than the galactocentric distance of the fountain. However, their
mass resolution (particle
mass $\simeq 1.5 \times 10^5 M_{\odot}$ in the highest resolution simulation) makes
their results more suitable for understanding the global behavior of the
HI in the Galaxy rather than the evolution of a single
superbubble/supershell. 
 In Tab. \ref{velocity} we report the main radial velocity for each fountains given by:
\begin{equation}
<v_{R}>=\frac{<\Delta R>}{<\Delta t>},
\end{equation}
where $\Delta R= R_{final}-R_{0}$ and $\Delta t=t_{final}+t_{orbit}$; the time $t_{orbit}$ is the time required for the cloud to return to the galactic disk once it leaves the supershell. Several chemical evolution studies have suggested that metallicity gradient disappears if the velocity of the radial flow is $|v_{f}|>$ 2 km s$^{-1}$ (see Matteucci 2001).

\begin{table}[h!]
\caption{Radial flow velocities (in km s$^{-1}$) in the solar neighborhood (model without drag) as a function of the radial throwing coordinate.}
\label{velocity}
\begin{center}
\begin{tabular}{cccc}
  \hline\hline

\noalign{\smallskip}
& 4 kpc   & 8 kpc &12 kpc\\
\noalign{\smallskip}

  \hline
\noalign{\smallskip}
  10 SNe& 0.6  &0.6&1.1\\
 50 SNe&1.6 &2.3&4.2\\
 100 SNe& 2.6 & 4.1& 6.6\\
 500 SNe& 4.5 & 13.2& 14.4\\

 \hline\hline
 \end{tabular}
\end{center}
\end{table}
For the model with viscous interaction we must know also the mass of the cloud because the drag terms depend on this quantity (see eq. \ref{tdrag2}). Referring to Fig. \ref{bubble} we assume that the part of the shell that could fragment and skip upwards is the mass included above the $s $ height where $\frac{dR}{dz}=0$. Over this height each point of the shell has a positive velocity component along the z axis. Therefore, we estimate that the masses of a gas cloud ejected by sequential SN events in the extra-planar halo are of the order of:
\begin{equation} 
M_{{cloud}_{R_0=4}}=2.17 \times 10^{5} M_{\odot},    
\label{1}
\end{equation}
\begin{equation}
M_{{cloud}_{R_0=8}}=1.24 \times 10^{5} M_{\odot},   
\label{2}
\end{equation}
\begin{equation}
M_{{cloud}_{R_0=12}}=0.84 \times 10^{5} M_{\odot}. 
\label{3}
\end{equation}
We note that the masses of the clouds are roughly 20\% of the initial supershell mass, therefore the total amount of the ejected metals in clouds are roughly 20 \% of the produced metals by the OB association, in agreement with the recent Melioli et al. (2008) hydrodynamical simulation results.

\begin{figure}
\label{8kpca}
\includegraphics[width=0.45\textwidth]{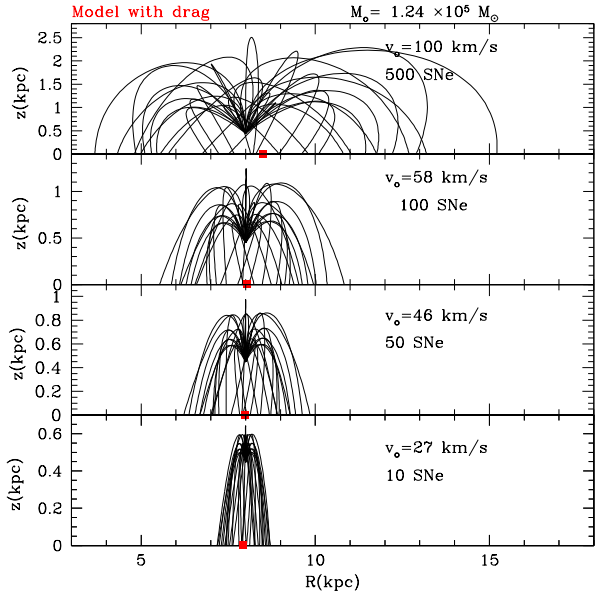}
\caption{ Galactic fountains in a model with drag with the initial radial coordinate $R_{0}$=8 kpc. In this case the orbits depend on the mass of the cloud ejected (indicated on the top of the panels). The natural effect of the viscous interaction is to decelerate the clouds.   }
\label{8kpca}
\end{figure}

\begin{figure}
\includegraphics[width=0.45\textwidth]{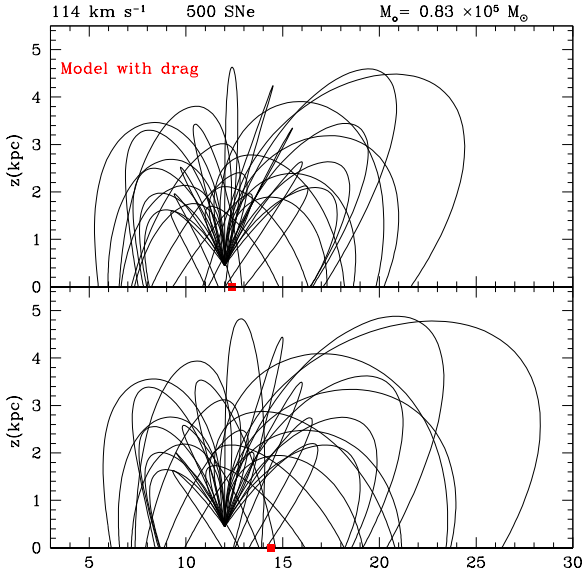}
\caption{Fountains which reach the maximum height in our model. }
\label{114misto}
\end{figure}

In Fig. \ref{8kpca} we show that for the range of initial velocities considered here, the effect of a viscous term in the motion equations is weak. The most evident effect is reported in Fig. \ref{114misto} where the hybrid model is compared with the purely ballistic one in the case of 500 sequential SN explosions at $R_0$= 12 kpc. The natural effect of a viscous interaction between the cloud and the extra-planar gas is to brake the motion of the clouds. Therefore, the average radial falling coordinate on the disk for the model with drag $<R_{final}>$= 12.37 kpc is much smaller than the one obtained with the purely ballistic model ($<R_{final}>$= 14.38 kpc) 

\begin{figure}[ht]
\begin{center}
\includegraphics[width=0.45\textwidth]{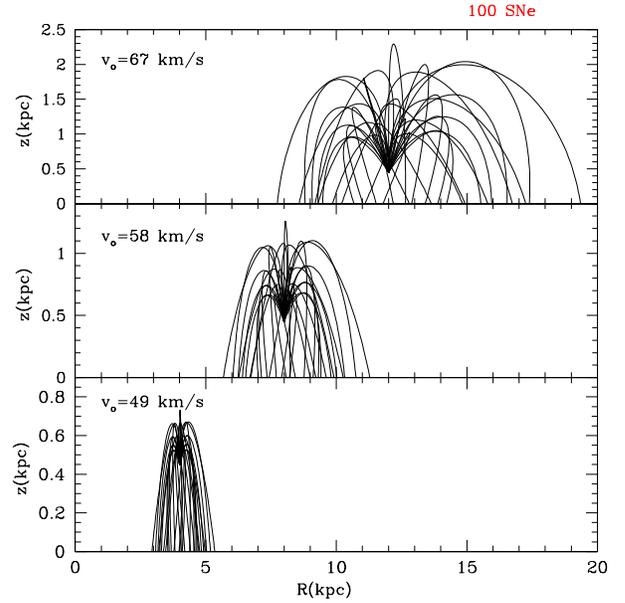} 
\caption{Fountains in the purely ballistic model driven by a sequential explosion of 100 SNe}  
\label{100SN}
\end{center}
\end{figure}

As shown in Tab. \ref{taze}, it is likely that the most realistic number of massive stars in OB associations in our Galaxy is about 100. In Fig. \ref{100SN} we have reproduced various fountains for the model without drag obtained with 100 sequential SN explosions respectively at 12, 8, 4 Kpc as radial radial coordinates. As shown before, different radial initial conditions imply different ejection velocities for the clouds. In particular, the throwing coordinate $R_0$=12 kpc is surrounded by a more tenuous disk, therefore the velocity is larger and the orbits wider. The highest z coordinate that a cloud could reach in this case is about $z$= 2.3 kpc.

\begin{table}[h!]
\caption{Massive stars in OB associations from P.T. de Zeeuw et al. (1998)}
\label{taze}
\begin{center}
\begin{tabular}{c ccccc}

\hline\hline
\noalign{\smallskip}
 OB association&$WR$&$O$&$B $& $A$   &Total\\
   \hline\hline
\noalign{\smallskip}
Upper Scorpius&&&49& 34 &83 \\

Upper Centaurus Lupus&&&66 &68    &134 \\

Lower Centaurus Crux&&&42 & 55  & 97 \\

Vela OB2&1&&81 & 5    &  87 \\

Trumpler 10&&&22 & 1   & 23 \\

Collinder 121&1&1&85&8 & 95\\

Perseus OB2&&&17 & 16   & 33 \\

$\alpha$ Perseus OB2&&&33 & 30   & 63 \\

Lacerta OB1&&1&35 & 46   & 82 \\

Cepheus OB2&&1&56 & 10   & 67 \\

\hline
\end{tabular}
\end{center}
\end{table}

\section{Comparison with HVCs and IVCs observations}\label{hvc}

In this section we compare our models with observational data concerning distance and chemical composition of some clouds in our Galaxy. We consider two cloud systems: Complex C (HVCs) and IV Arch (IVC).


\subsection{Complex C}
In the work of Wakker et al. (2007) it was reported the first successful detection of interstellar $Ca$\,{\sc ii} $H$ and $K$ absorption from
HVC complex~C. They concluded that complex C is located at Galactocentric radius $<$14 kpc, and lies high above the Galactic plane ($z$=3--9 kpc).

 Integrating the  $H$\,{\sc i} column density across the cloud, it was estimated $M$( $H$\,{\sc i}) as 
0.7-6.0 $\times 10^{6} M_{\odot}$. As we show in eqs. (\ref{1}), (\ref{2}), (\ref{3}), the sizes of our ejected clouds are considerably smaller. Even if we consider the case with the superbubble given by  500 SN explosions with $R_0$=12 kpc, the highest $z$  we can reach with this event is about $z  \simeq 4.4$ kpc and at this height our cloud velocities are roughly equal to  zero. In the Tab. 1 of Wakker et al. (2007) are reported the velocities (relative to the LSR) of complex C in the direction of different line of sight, and in average the complex C velocity is -114 km/s. Therefore, the kinematical data of Complex C are inconsistent with our results.

 A key point for understanding the origin of this HVC is considering its chemical composition. Because oxygen is not significantly depleted 
onto dust grains (Meyer et al.\,1998) and the $O$\,{\sc i}/$H$\,{\sc i} ratio is not
altered by ionization effects, we have that ({$O$}{\sc i}/{$H$}{\sc i}) $\approx$ ($O$/$H$). In the work of Richter et al. (2001) the solar values for the abundances are given by Anders \& Grevesse (1989) and  Grevesse \& Noels (1993). For the oxygen they obtained:
\begin{equation}
\label{O/H}
[O/H] \simeq  [O\,\mbox{\,{\sc i}}/H \mbox{\,{\sc i}}]= -1.03^{+0.37}_{-0.31}.
\end{equation}
For the iron abundance in the Complex C it was obtained:
\begin{equation}
[Fe/H]\simeq [(Fe   \mbox{\,{\sc ii}}+Fe\mbox{\,{\sc iii}})/H\mbox{\,{\sc i}}].
= -1.15,
\end{equation}
Based on the data from $Fe\,\mbox{\,{\sc ii}}$ they obtained
$[Fe/H]=-1.27^{+0.20}_{-0.14}$. Therefore a low iron abundance in
Complex C could be a result of depletion of iron into $Fe$ rich dust
(Savage \& Sembach 1996). A more careful discussion about the
possible effect of dust is in Sect. 4.3.  Hence the relative
abundance is $[O/Fe]=0.12$. In our model a similar result is obtained
in the case of 500 SNe with initial radial throwing coordinate fixed
at $R_0=12$ kpc and, as disk initial gas metallicity, $Z= 0.1 \times
Z_{\odot}$.  As we have seen, the most likely metallicity of the disk
at $R_0$= 12 kpc at the present time is 0.74 $Z_{\odot}$, which
is not consistent with the total metallicity of Complex C (eqs. 50 and
51).  We obtain with our model $[O/Fe]=0.03$ dex in the case of 500
SNe with throwing coordinate $R_0=12$ kpc and metallicity fixed at
0.74 $Z_{\odot}$ (see Fig. \ref{zreal}).  Such an oxygen
overabundance, although not impossible, is quite unlikely.  Given the
inconsistency of the kinematical data with our predictions, we can
rule out a Galactic origin for the Complex C HVC.

 \subsection{IV ARCH}

Richter et al. (2001) suggested that IV Arch, given its nearly solar
abundance and its $z$-height bracket of 0.8-1.5 kpc, has its origin in
the Milky Way disk, probably as part of a Galactic fountain. These
heights are indeed consistent with our results, and also the
velocities of IVCs are easily understandable in the framework of
galactic fountains. The abundances measured by Richter et al. (2001)
are: $[O/H]=-0.01^{+0.35}_{-0.27}$ and $[Fe/H]=-0.26^{+0.19}_{-0.15}$,
therefore:

\begin{equation}
\label{IV}
[O/Fe] = [O/H]-[Fe/H]= 0.25 \, \mbox{dex}.
\end{equation}
If we assume a Galactic fountain origin, our model yields
$[O/Fe]=0.24$ dex in the case of 500 SN explosions, $R_0$= 8 kpc and
metallicity $Z=0.01 \times Z_{\odot}$, and $[O/Fe]=0.26$ dex in the
case of 100 SN explosions, $R_0$= 12 kpc and metallicity $Z=0.01
\times Z_{\odot}$.  As we have seen, it is quite unlikely that the
initial metallicity of an OB association can be nowadays as low as
0.01 $Z_{\odot}$ and this low initial metallicity would be
inconsistent with the nearly solar abundance measured in IV
Arch. Therefore, although the kinematics properties of IV Arch are
in agreement with our results, its large $[O/Fe]$ casts some doubt
about the Galactic origin of this IVC.

\subsection{The effect of dust} 

In the work of Lu et al (1998) it was shown how important is the
depletion of metals (in particular $Fe$) in the ISM into dust. Since
this can affect out results about $[O/Fe]$ ratios, here we estimate
the time scales of the destruction and accretion of dust in a
superbubble.  Referring to the work of Calura et al. (2008) we define,
for a given element $i$ the destruction time-scale $\tau_{destr}$ for
the dust in the superbubble due to SN shocks as:
\begin{equation}
\tau_{destr, i}=(\epsilon M_{SNR})^{-1} \cdot \frac{\sigma_{ISM}}{R_{SN}},
\end{equation}
where $R_{SN}$ is the total SN rate, $M_{SNR}$ is the mass of the
interstellar gas swept up by the SN remnant and $\epsilon$ is the
destruction efficiency in a three-phase medium. We consider for
$\sigma_{ISM}$, the surface gas density of the ISM, the value of $10
M_{\odot}$ pc$^{-2}$ and $\epsilon M_{SNR} \simeq 10^5 M_{\odot}$. For
the SN rate we consider the typical case of the OB association with
100 SNe.  We assume as radius of the OB association the conservative
value of 100 pc. In the end we obtain that $\tau_{destr, i}\simeq$ 0.7
Myr. This timescale is considerably smaller than the time necessary
for the formation of a RT instable supershell in our model (See
sect. \ref{clouds} ). If the number of SN is smaller, $\tau_{destr}$
is larger but it remains considerably smaller than $t_{final}$.  Given
the complexity of the interaction between SN shocks and dust (Draine \&
Salpeter 1979; Jones et al. 1996) this estimate is necessarily
simplified but it leads us to the suggestion that only a small amount
of dust can survive in the supershell swept up by the OB association,
therefore only a small fraction of metals in the clouds driven by the
galactic fountains is locked into dust grains. Can this amount
significantly increase by dust accretion during the journey of the
cloud? To answer this question we have to compare the dynamical
timescale of our clouds with the typical dust accretion timescale.

Following the work of Calura et al. (2008) let $X_{d}$ be the
abundance by mass of the dust and $\sigma$ the ISM fraction at the
time $t$, the quantity $\sigma_{dust}= X_{d} \cdot \sigma$ represents
the normalized mass density of the dust at the time \emph{t}.

The accretion timescale $\tau_{accr}$ is given by:

\begin{equation}
\tau_{accr}=\tau_{0}/(1 - X_d) 
\label{accr_t},
\end{equation}
for $\tau_{0}$ typical values range between 50 Myr and 200 Myr (see
Dwek 1998). The integration of the equation for the temporal evolution
of $X_{d}$ with the initial condition $X_{d,0} \simeq 0$ and $X_d\ll
1$ gives:

\begin{equation}
t  \simeq \tau_{0,i}\left[ \ln(X_d/X_{d,0}) - \ln(1) \right] 
\end{equation}
We want to know how much time does it take to increase the initially
small amount of dust up to a non-negligible value.  If we calculate
for instance the time necessary for accreting 10 times the initial
dust fraction (e.g. $X_d/X_{d,0}=10$), we get:

\begin{equation}
  t\simeq 2.3 \tau_{0,i}.
\end{equation}
This time is greater than the average time of the cloud orbits,
therefore we can conclude that the depletion of metals into dust does
not have a important role in the supershell evolution.

\section{Conclusions}\label{conclusions}

In this paper we have studied the evolution of a supershell powered by the energetic feedback of a typical Galactic OB association at various Galactocentric radii. Based on the Kompaneets (1960) approximation, we have found analytical solutions of the temporal evolution of the supershell and we have established criteria for its fragmentation, which can create clouds that are thrown out of the disk. Given the self-similar behavior of the Kompaneets solutions, the clouds are formed at the same scale height ($\sim$ 450 pc), irrespective of the number of SNe in the OB association or of the Galactocentric radius at which the OB association are. Assuming that the ejecta of the dying stars of the OB association instantaneously mix with the supershell, we are able to calculate the chemical composition of the clouds and in particular their [$O/Fe$].

We have considered four different OB associations (containing 10, 50, 100 and 500 SNe, respectively) and three different initial throwing coordinates (4, 8 and 12 Kpc, respectively). Once the clouds are formed and can leave the disk, we follow their orbits either assuming a purely ballistic model, or introducing a viscous force acting between the cloud and the surrounding hot halo gas. Our main conclusions can be summarized as follows:

\begin{itemize}

\item If the initial metallicity of the OB association is solar, the pollution from the dying stars has a negligible effect on the chemical composition of the clouds. In particular, the [$O/Fe$] abundance ratio reaches at most $\sim$ 0.025 in a model in which the throwing coordinate is R$_0$ = 12 kpc. Only starting from very low metallicities (less than 1/100 Z$_\odot$) it is possible to produce a significant enrichment of $\alpha$-elements.

\item Both in the ballistic and in the viscous interaction models the maximum height reached by the clouds is not very large. Only for OB associations composed of 500 SNe it is possible to throw clouds up to heights larger than 2 kpc above the plane of the Galaxy.

\item The range of the cloud orbits is also quite small. The clouds are generally directed outwards but the average landing coordinates differ from the throwing coordinates by $\sim$ 1 kpc at most. Only for a throwing coordinate of 12 kpc and an OB association made of 500 SNe the ballistic model predicts a landing coordinate $\sim$ 2 kpc larger than the throwing one. 

\item Models including a viscous interaction between clouds and the extra-planar gas predict smaller ranges of the cloud orbits. Indeed, the drag experienced by the cloud brakes it and therefore it shortens its journey above the Galactic plane.

\item The HVC Complex C has a mass, velocity and inferred height above the plane inconsistent with the results of our models. Its oxygen overabundance ([$O/Fe$]=0.12) can be reproduced only if we assume a large OB association with metallicity 0.1 Z$_\odot$ or smaller, therefore its Galactic origin cannot be completely ruled out on the basis of its chemical composition alone, but the kinematical data suggest a different formation mechanism. The IVC Arch IV has instead velocities and heights above the plane easily reproduced by our models, but its [$O/Fe$] =0.25 is much larger than the one of Complex C and it can be explained only by assuming initial metallicities smaller than 1/100 Z$_\odot$ which are unlikely at the present time for the Galactic disk. Therefore, it is unlikely that the two studied clouds are originated in a Galactic fountain motion.

\end{itemize}


\begin{acknowledgements}

We thank the referee for the enlightening suggestions.  We are grateful to A. D'Ercole for having kindly provided a ballistic galactic fountains code. We also thank F. Calura, G. Cescutti for many useful discussions.

\end{acknowledgements}

\section{Online Material}
In the following tables we summarize all our results concerning the abundances of oxygen and iron calculated with our model for the initial throwing radial coordinates $R_0=4, 8, 12$ kpc and for metallicities: $Z_{\odot}$, $0.1 \times Z_{\odot}$, $0.01 \times Z_{\odot}$, $10^{-4} \times Z_{\odot}$.

\begin{table*}[htp]

\caption{$Z=Z_{\odot}$ 8 kpc}
\label{solar8kpc}
\begin{center}
\begin{tabular}{cccccccccc}
  \hline\hline
\noalign{\smallskip}

& & & & & &\\
 Number &$M_{*Fe_{56}} $&$M_{*O_{16}} $& $M_{*ej}$   &$M_{Shell Fe_{56}}$&$M_{Shell O_{16}}$&$M_{tot}$& $X_{*Fe_{56}}$& $X_{*O_{16}}$&$[O/Fe]$\\
  of SNe & & & & & &\\
& & & & & &\\
\hline
& & & & & &\\
10&0.95 &10.28 &$ 1.16\times 10^{2}$ &678.40& $5.56\times 10^{3}$&$57.91\times 10^{4}$ &$1.17\times 10^{-3}$& $9.60\times 10^{-3}$&$1.97\times 10^{-4}$\\

& & & & & &\\
50&3.95 &49.89 &$ 4.77\times 10^{2}$    &681.40& $5.60\times 10^{3}$ &$57.94\times 10^{4}$ &$1.17\times 10^{-3}$& $9.67\times 10^{-3}$&$1.36\times 10^{-3}$\\

& & & & & &\\
100&6.99 &97.55 &$ 8.52\times 10^{2}$   &684.44& $5.69\times 10^{3}$ &$57.98\times 10^{4}$&$1.18\times 10^{-3}$& $9.75\times 10^{-3}$&$3.11\times 10^{-3}$\\

& & & & & &\\  
500&24.69 &437.46&  $ 3.18\times 10^{3}$    &702.14&$5.99\times 10^{3}$&$58.22\times 10^{4}$ &$1.21\times 10^{-3}$& $1.03\times 10^{-2}$&$1.74\times 10^{-2}$\\

& & & & &&\\
\hline

\end{tabular}
\end{center}

\end{table*}

\begin{table*}[htp]
\caption{$Z=0.1 \times Z_{\odot}$  8 kpc}

\label{01solar}
\begin{center}
\begin{tabular}{cccccccccc}
  \hline\hline
\noalign{\smallskip}

& & & & & &\\
 Number &$M_{*Fe_{56}} $&$M_{*O_{16}} $& $M_{*ej}$   &$M_{Shell Fe_{56}}$&$M_{Shell O_{16}}$&$M_{tot}$& $X_{*Fe_{56}}$& $X_{*O_{16}}$&$[O/Fe]$\\
  of SNe & & & & & &\\
& & & & &&\\
\hline\
& & & & & &\\
10&0.95 &9.33 &$ 1.14\times 10^{2}$ &68.70& $ 5.64\times 10^{2}$&$57.91\times 10^{4}$ &$1.19\times 10^{-4}$&$9.75\times 10^{-4}$ &$1.17\times 10^{-3}$\\
& & & & &&\\

50&3.95 & 45.48&$ 4.81\times 10^{2}$    &71.70&$ 6.01\times 10^{2}$ &$57.95\times 10^{4}$ &$1.24\times 10^{-4}$& $1.04\times 10^{-3}$&$9.55\times 10^{-3}$\\
& & & & & &\\

100&6.99&88.97 &$ 8.59\times 10^{2}$   &75.57&$ 6.44\times 10^{2}$ &$57.99\times 10^{4}$&$1.30\times 10^{-4}$& $1.11\times 10^{-3}$&$1.71\times 10^{-2}$\\
& & & & & &\\

500&24.69 & 401.60 &$ 3.21\times 10^{3}$    &92.43& $ 9.57\times 10^{2}$&$58.22\times 10^{4}$ &$1.59\times 10^{-4}$& $1.64\times 10^{-3}$ &$1.01\times 10^{-1}$\\

& & & & &&\\
\hline

\end{tabular}
\end{center}

\end{table*}

\newpage

\begin{table*}[htp]
\caption{$Z=0.01 \times Z_{\odot}$ 8 kpc}
\label{0.01solar}

\begin{center}
\begin{tabular}{cccccccccc}
  \hline\hline
\noalign{\smallskip}

& & & & &&\\
 Number &$M_{*Fe_{56}} $&$M_{*O_{16}} $& $M_{*ej}$   &$M_{Shell Fe_{56}}$&$M_{Shell O_{16}}$&$M_{tot}$& $X_{*Fe_{56}}$& $X_{*O_{16}}$&$[O/Fe]$\\
 of SNe  & & & & &&\\
& & & & &&\\
\hline
& & & & & &\\
10&0.95 & 8.96&$ 1.14\times 10^{2}$ &7.72&  64.50&$57.91\times 10^{4}$ &$1.33\times 10^{-5}$&$1.11\times 10^{-4}$ &$8.29\times 10^{-3}$\\
& & & & &&\\

50&3.95 &43.67 &$ 4.82\times 10^{2}$    &10.73& 99.20&$57.95\times 10^{4}$ &$1.85\times 10^{-5}$& $1.71\times 10^{-4}$&$5.23\times 10^{-2}$\\
& & & & & &\\

100&6.99&85.72 &$ 8.61\times 10^{2}$   &13.77& $ 1.41\times 10^{2}$&$57.99\times 10^{4}$&$2.37\times 10^{-5}$& $2.43\times 10^{-4}$&$9.74\times 10^{-2}$\\
& & & & &&\\

500&24.69 & 394.21 &$ 3.23\times 10^{3}$    &31.46&$ 4.50\times 10^{2}$ &$58.22\times 10^{4}$ &$5.40\times 10^{-5}$&$7.72\times 10^{-4}$ &$2.41\times 10^{-1}$\\

& & & & &&\\
\hline

\end{tabular}
\end{center}

\end{table*}

\begin{table*}[htp]
\caption{$Z=10^{-4} \times Z_{\odot}$ 8 kpc }
\label{ultima}

\begin{center}
\begin{tabular}{cccccccccc}
  \hline\hline
\noalign{\smallskip}

& & & & & &\\
 Number &$M_{*Fe_{56}} $&$M_{*O_{16}} $& $M_{*ej}$   &$M_{Shell Fe_{56}}$&$M_{Shell O_{16}}$&$M_{tot}$& $X_{*Fe_{56}}$& $X_{*O_{16}}$&$[O/Fe]$\\
 of SNe  & & & & &&\\
& & & & &&\\
\hline
& & & & &&\\
10&0.95 & 8.55&$ 1.14\times 10^{2}$ &1.02&9.10 &$57.91\times 10^{4}$ &$2.09\times 10^{-6}$&$ 1.57\times 10^{-5}$ &$ 3.68\times 10^{-2}$\\

& & & & &&\\
50&3.95 & 41.54&$ 4.81\times 10^{2}$    &4.02&42.10 &$57.95\times 10^{4}$ &$7.26\times 10^{-6}$& $ 7.26\times 10^{-5}$&$ 1.06\times 10^{-1}$\\

& & & & & &\\
100&6.99&81.58 &$ 8.60\times 10^{2}$   &7.06& 82.14&$57.99\times 10^{4}$&$1.22\times 10^{-5}$&$ 1.42\times 10^{-4}$ &$ 1.52\times 10^{-1}$\\

& & & & &&\\
500&24.69 &377.66&$ 3.23\times 10^{3}$    &24.75& $3.78\times 10^{2}$&$58.22\times 10^{4}$ &$4.25\times 10^{-5}$& $ 6.52\times 10^{-4}$&$2.70\times 10^{-1}$\\

& & & & &&\\
\hline

\end{tabular}
\end{center}

\end{table*}

\begin{table*}[htp]

\caption{$Z=Z_{\odot}$ 12 kpc}
\label{solar 12kpc}
\begin{center}
\begin{tabular}{cccccccccc}
  \hline\hline
\noalign{\smallskip}

& & & & & &\\
 Number &$M_{*Fe_{56}} $&$M_{*O_{16}} $& $M_{*ej}$   &$M_{Shell Fe_{56}}$&$M_{Shell O_{16}}$&$M_{tot}$& $X_{*Fe_{56}}$& $X_{*O_{16}}$&$[O/Fe]$\\
  of SNe & & & & & &\\
& & & & & &\\
\hline
& & & & & &\\
10&0.92 &10.23 &$ 1.13\times 10^{2}$ &456.70& $3.75\times 10^{3}$&$38.97\times 10^{4}$ &$1.17\times 10^{-3}$& $9.61\times 10^{-3}$&$3.13\times 10^{-4}$\\

& & & & & &\\
50&3.74 &49.28 &$ 4.47\times 10^{2}$    &459.52& $3.78\times 10^{3}$ &$39.00\times 10^{4}$ &$1.18\times 10^{-3}$& $9.70\times 10^{-3}$&$2.14\times 10^{-3}$\\

& & & & & &\\
100&6.35 &95.73 &$ 7.95\times 10^{2}$   &462.13& $3.83\times 10^{3}$ &$39.03\times 10^{4}$&$1.18\times 10^{-3}$& $9.81\times 10^{-3}$&$4.98\times 10^{-3}$\\

& & & & & &\\
500&23.27 &421.95&  $ 2.95\times 10^{3}$    &479.05&$4.16\times 10^{3}$&$39.25\times 10^{4}$ &$1.22\times 10^{-3}$& $1.06\times 10^{-2}$&$2.48\times 10^{-2}$\\

& & & & &&\\
\hline

\end{tabular}
\end{center}

\end{table*}

\begin{table*}[htp]

\caption{$Z=0.1 \times Z_{\odot}$  12 kpc}
\label{0.1 12kpc}
\begin{center}
\begin{tabular}{cccccccccc}
  \hline\hline
\noalign{\smallskip}

& & & & & &\\
 Number &$M_{*Fe_{56}} $&$M_{*O_{16}} $& $M_{*ej}$   &$M_{Shell Fe_{56}}$&$M_{Shell O_{16}}$&$M_{tot}$& $X_{*Fe_{56}}$& $X_{*O_{16}}$&$[O/Fe]$\\
  of SNe & & & & & &\\
& & & & & &\\
\hline
& & & & & &\\
10&0.92 &9.29 &$ 1.11\times 10^{2}$ &46.50& $3.83\times 10^{2}$&$38.97\times 10^{4}$ &$1.19\times 10^{-4}$& $9.82\times 10^{-4}$&$1.98\times 10^{-3}$\\

& & & & & &\\
50&3.74 &45.00 &$ 4.51\times 10^{2}$    &49.32& $4.18\times 10^{2}$ &$39.00\times 10^{4}$ &$1.26\times 10^{-4}$& $1.07\times 10^{-3}$&$1.51\times 10^{-2}$\\

& & & & & &\\
100&6.35 &87.27 &$ 8.02\times 10^{2}$   &51.92& $4.61\times 10^{2}$ &$39.03\times 10^{4}$&$1.33\times 10^{-4}$& $1.18\times 10^{-3}$&$3.46\times 10^{-2}$\\

& & & & & &\\
500&23.27 &388.11&  $ 2.97\times 10^{3}$    &68.85&$7.62\times 10^{2}$&$39.25\times 10^{4}$ &$1.75\times 10^{-4}$& $1.94\times 10^{-3}$&$1.30\times 10^{-1}$\\

& & & & &&\\
\hline

\end{tabular}
\end{center}

\end{table*}

\begin{table*}[htp]

\caption{$Z=0.01 \times Z_{\odot}$ 12 kpc}
\label{0.01 12kpc}
\begin{center}
\begin{tabular}{cccccccccc}
  \hline\hline
\noalign{\smallskip}

& & & & & &\\
 Number &$M_{*Fe_{56}} $&$M_{*O_{16}} $& $M_{*ej}$   &$M_{Shell Fe_{56}}$&$M_{Shell O_{16}}$&$M_{tot}$& $X_{*Fe_{56}}$& $X_{*O_{16}}$&$[O/Fe]$\\
  of SNe & & & & & &\\
& & & & & &\\
\hline
& & & & & &\\
10&0.92 &8.93 &$ 1.11\times 10^{2}$ &5.48& 46.29&$38.97\times 10^{4}$ &$1.41\times 10^{-5}$& $1.19\times 10^{-4}$&$1.31\times 10^{-2}$\\

& & & & & &\\
50&3.74 &43.24 &$ 4.52\times 10^{2}$    &8.30& 80.60 &$39.00\times 10^{4}$ &$2.13\times 10^{-5}$& $2.07\times 10^{-4}$&$7.36\times 10^{-2}$\\

& & & & & &\\
100&6.35 &84.64 &$ 8.06\times 10^{2}$   &10.90& 121.82 &$39.04\times 10^{4}$&$2.79\times 10^{-5}$& $3.12\times 10^{-4}$&$1.35\times 10^{-1}$\\

& & & & & &\\
500&23.27 &381.99&  $ 3.00\times 10^{3}$    &27.83&419.35&$39.25\times 10^{4}$ &$7.09\times 10^{-5}$& $1.07\times 10^{-3}$&$2.64\times 10^{-1}$\\

& & & & &&\\
\hline

\end{tabular}
\end{center}

\end{table*}

\begin{table*}[htp]

\caption{$Z=10^{-4} \times Z_{\odot}$ 12 kpc }
\label{0.01 12kpc}
\begin{center}
\begin{tabular}{cccccccccc}
  \hline\hline
\noalign{\smallskip}

& & & & & &\\
 Number &$M_{*Fe_{56}} $&$M_{*O_{16}} $& $M_{*ej}$   &$M_{Shell Fe_{56}}$&$M_{Shell O_{16}}$&$M_{tot}$& $X_{*Fe_{56}}$& $X_{*O_{16}}$&$[O/Fe]$\\
  of SNe & & & & & &\\
& & & & & &\\
\hline
& & & & & &\\
10&0.92 &8.51 &$ 1.11\times 10^{2}$ &0.97& 8.88&$38.97\times 10^{4}$ &$2.49\times 10^{-6}$& $2.28\times 10^{-5}$&$4.80\times 10^{-2}$\\

& & & & & &\\
50&3.74 &41.13 &$ 4.52\times 10^{2}$    &3.79& 41.5 &$39.00\times 10^{4}$ &$9.72\times 10^{-6}$& $1.06\times 10^{-4}$&$1.26\times 10^{-1}$\\

& & & & & &\\
100&6.35 &80.47 &$ 8.05\times 10^{2}$   &6.39& 80.84 &$39.04\times 10^{4}$&$1.64\times 10^{-5}$& $2.07\times 10^{-4}$&$1.88\times 10^{-1}$\\

& & & & & &\\
500&23.27 &367.43&  $ 2.99\times 10^{3}$    &23.31&367.81&$39.25\times 10^{4}$ &$5.94\times 10^{-5}$& $9.37\times 10^{-4}$&$2.84\times 10^{-1}$\\

& & & & &&\\
\hline

\end{tabular}
\end{center}

\end{table*}

\begin{table*}[htp]

\caption{$Z=Z_{\odot}$ 4 kpc}
\label{solar}
\begin{center}
\begin{tabular}{cccccccccc}
  \hline\hline
\noalign{\smallskip}

& & & & & &\\
 Number &$M_{*Fe_{56}} $&$M_{*O_{16}} $& $M_{*ej}$   &$M_{Shell Fe_{56}}$&$M_{Shell O_{16}}$&$M_{tot}$& $X_{*Fe_{56}}$& $X_{*O_{16}}$&$[O/Fe]$\\
  of SNe & & & & & &\\
& & & & & &\\
\hline
& & & & & &\\
10&0.97 &10.32 &$ 1.19\times 10^{2}$ &1179.14& $9.67\times 10^{3}$&$100.71\times 10^{4}$ &$1.17\times 10^{-3}$& $9.60\times 10^{-3}$&$1.04\times 10^{-4}$\\

& & & & & &\\
50&4.26 &50.57 &$ 5.20\times 10^{2}$    &1182.42& $9.71\times 10^{3}$ &$100.75\times 10^{4}$ &$1.17\times 10^{-3}$& $9.63\times 10^{-3}$&$7.02\times 10^{-4}$\\

& & & & & &\\
100&7.79 &99.37 &$ 9.33\times 10^{2}$   &1185.95& $9.76\times 10^{3}$ &$100.79\times 10^{4}$&$1.18\times 10^{-3}$& $9.69\times 10^{-3}$&$1.58\times 10^{-3}$\\

& & & & & &\\
500&27.36 &457.84&  $ 3.53\times 10^{3}$    &1205.52&$10.11\times 10^{3}$&$101.05\times 10^{4}$ &$1.19\times 10^{-3}$& $1.00\times 10^{-2}$&$1.01\times 10^{-2}$\\

& & & & &&\\
\hline

\end{tabular}
\end{center}

\end{table*}

\begin{table*}[htp]

\caption{$Z=0.1 \times Z_{\odot}$  4 kpc}
\label{solar}
\begin{center}
\begin{tabular}{cccccccccc}
  \hline\hline
\noalign{\smallskip}

& & & & & &\\
 Number &$M_{*Fe_{56}} $&$M_{*O_{16}} $& $M_{*ej}$   &$M_{Shell Fe_{56}}$&$M_{Shell O_{16}}$&$M_{tot}$& $X_{*Fe_{56}}$& $X_{*O_{16}}$&$[O/Fe]$\\
  of SNe & & & & & &\\
& & & & & &\\
\hline
& & & & & &\\
10&0.97 &9.36 &$ 1.16\times 10^{2}$ &118.79& $9.75\times 10^{2}$&$100.71\times 10^{4}$ &$1.18\times 10^{-4}$& $9.68\times 10^{-4}$&$6.17\times 10^{-4}$\\

& & & & & &\\
50&4.26 &45.99 &$ 5.19\times 10^{2}$    &122.08& $10.11\times 10^{2}$ &$100.75\times 10^{4}$ &$1.21\times 10^{-4}$& $1.00\times 10^{-3}$&$4.77\times 10^{-3}$\\

& & & & & &\\
100&7.79 &90.67 &$ 9.41\times 10^{2}$   &125.61& $10.56\times 10^{2}$ &$100.79\times 10^{4}$&$1.25\times 10^{-4}$& $1.05\times 10^{-3}$&$1.11\times 10^{-2}$\\

& & & & & &\\
500&27.36 &418.89&  $ 3.56\times 10^{3}$    &145.17&$13.84\times 10^{2}$&$101.05\times 10^{4}$ &$1.44\times 10^{-4}$& $1.37\times 10^{-3}$&$6.58\times 10^{-2}$\\

& & & & &&\\
\hline

\end{tabular}
\end{center}

\end{table*}

\begin{table*}[htp]

\caption{$Z=0.01 \times Z_{\odot}$ 4 kpc}
\label{solar}
\begin{center}
\begin{tabular}{cccccccccc}
  \hline\hline
\noalign{\smallskip}

& & & & & &\\
 Number &$M_{*Fe_{56}} $&$M_{*O_{16}} $& $M_{*ej}$   &$M_{Shell Fe_{56}}$&$M_{Shell O_{16}}$&$M_{tot}$& $X_{*Fe_{56}}$& $X_{*O_{16}}$&$[O/Fe]$\\
  of SNe & & & & & &\\
& & & & & &\\
\hline
& & & & & &\\
10&0.97 &9.00 &$ 1.16\times 10^{2}$ &12.76& $1.05\times 10^{2}$&$100.71\times 10^{4}$ &$1.27\times 10^{-5}$& $1.05\times 10^{-4}$&$4.06\times 10^{-3}$\\

& & & & & &\\
50&4.26 &44.18 &$ 5.20\times 10^{2}$    &16.04& $1.41\times 10^{2}$ &$100.75\times 10^{4}$ &$1.59\times 10^{-5}$& $1.40\times 10^{-4}$&$2.96\times 10^{-2}$\\

& & & & & &\\
100&7.79 &87.06 &$ 9.43\times 10^{2}$   &19.57& $1.84\times 10^{2}$ &$100.79\times 10^{4}$&$1.94\times 10^{-5}$& $1.82\times 10^{-4}$&$5.87\times 10^{-2}$\\

& & & & & &\\
500&27.36 &408.93&  $ 3.58\times 10^{3}$    &39.14&$5.05\times 10^{2}$&$101.06\times 10^{4}$ &$3.87\times 10^{-5}$& $5.00\times 10^{-4}$&$1.97\times 10^{-1}$\\

& & & & &&\\
\hline

\end{tabular}
\end{center}

\end{table*}

\begin{table*}[htp]

\caption{$Z=10^{-4} \times Z_{\odot}$ 4 kpc }
\label{4kpc104}
\begin{center}
\begin{tabular}{cccccccccc}
  \hline\hline
\noalign{\smallskip}

& & & & & &\\
 Number &$M_{*Fe_{56}} $&$M_{*O_{16}} $& $M_{*ej}$   &$M_{Shell Fe_{56}}$&$M_{Shell O_{16}}$&$M_{tot}$& $X_{*Fe_{56}}$& $X_{*O_{16}}$&$[O/Fe]$\\
  of SNe & & & & & &\\
& & & & & &\\
\hline
& & & & & &\\
10&0.97 &8.58 &$ 1.16\times 10^{2}$ &1.09& 9.55&$100.71\times 10^{4}$ &$1.08\times 10^{-6}$& $9.48\times 10^{-6}$&$2.89\times 10^{-2}$\\

& & & & & &\\
50&4.26 &42.06 &$ 5.20\times 10^{2}$    &4.38& 43.03 &$100.75\times 10^{4}$ &$4.35\times 10^{-6}$& $4.27\times 10^{-5}$&$7.86\times 10^{-2}$\\

& & & & & &\\
100&7.79 &82.80 &$ 9.42\times 10^{2}$   &7.91& 83.77 &$100.79\times 10^{4}$&$7.85\times 10^{-6}$& $8.31\times 10^{-5}$&$1.11\times 10^{-1}$\\

& & & & & &\\
500&27.36 &390.33&  $ 3.58\times 10^{3}$    &27.47&391.30&$101.05\times 10^{4}$ &$2.71\times 10^{-5}$& $3.87\times 10^{-4}$&$2.40\times 10^{-1}$\\

& & & & &&\\
\hline

\end{tabular}
\end{center}

\end{table*}


\begin{thebibliography}{99}
\bibitem[\protect\citeauthoryear{Anders, E., \& Grevesse}{1989}]{anders89} Anders, E., \& Grevesse, N. 1989, Geochim. Cosmochim. Acta 53, 197

\bibitem[\protect\citeauthoryear{Andrievsky et al.}{2002}]{b10}
  Andrievsky S.M., Bersier D., Kovtyukh V. V. et al
  2002a, A\&A 384, 140


\bibitem[\protect\citeauthoryear{Andrievsky et al.}{2002}]{b20}
  Andrievsky S.M., Kovtyukh V. V., Luck R.E. et al
  2002b, A\&A 381, 32

\bibitem[\protect\citeauthoryear{Andrievsky et al.}{2002}]{b30}
  Andrievsky S. M., Kovtyukh V. V., Luck R.E. et al
  2002c, A\&A 392, 491

\bibitem[\protect\citeauthoryear{Andrievsky et al.}{2004}]{b40}
  Andrievsky S. M., Luck R. E., Martin P. et al
  2004, A\&A 413, 159

\bibitem[\protect\citeauthoryear{Asplund et al.}{2005}]{b55}
  Asplund, M., Grevesse, N., Sauval, A. J. 2005, in ``Cosmic Abundances as Records of Stellar Evolution and Nucleosynthesis'', eds T. G. Barnes III, F. N., Bash, ASP Conf. Ser. 336, 25 



\bibitem[\protect\citeauthoryear{Booth \& Theuns}{2007}]{b78}
  Booth, C. M., Theuns, T. 2007, MNRAS, 381, 89



\bibitem[\protect\citeauthoryear{Barnab\`e et al.}{2006}]{b01} Barnab\`e, M., Ciotti, L., Fraternali, F.,  Sancisi, R., 2006, A\&A, 446, 61
\bibitem[\protect\citeauthoryear{Basu et al.}{1999}]{b02} Basu, S., Johnstone, D., Martin, P. G., 1999, ApJ, 516, 843
\bibitem[\protect\citeauthoryear{Blitz et al.}{1999}]{b03} Blitz, L., Spergel, D. N., Teuben, P. J., Hartmann, D., Burton, W. B. 1999, ApJ 514, 818

\bibitem[\protect\citeauthoryear{Bluhm et al.}{2001}]{b06} Bluhm, H., de Boer, K., Marggraf, O., Richter, P. 2001, A\&A, 367, 299

\bibitem[\protect\citeauthoryear{Boomsma et al.}{2005}]{b04} Boomsma, R., Oosterloo, T., Fraternali, F., van der Hulst, T., Sancisi, R.  2005, in ``Extra-planar Gas''Conference, ASP Conf. Series, ed. R.~Braun, vol. 331, p. 247




\bibitem[\protect\citeauthoryear{Bregman et al.}{1980}]{b05} Bregman, J. N. 1980, ApJ, 365, 544


\bibitem[\protect\citeauthoryear{Calura et al.}{2008}]{b57} Calura, F.; Pipino, A.; Matteucci, F., 2008, A\&A, 479, 669
\bibitem[\protect\citeauthoryear{Cescutti et al.}{2006}]{b99} Cescutti, G., Matteucci, F., Francois, P., Chiappini, C. 2006, A\&A, 462, 943
\bibitem[\protect\citeauthoryear{Charlton et al.}{1989}]{b07} Charlton, J. C. \& Salpeter, E.E. 1989, ApJ, 346,101

\bibitem[\protect\citeauthoryear{Chiappini et al.}{2001}]{b67} Chiappini, C., Matteucci, F., Romano, D., 2001, ApJ, 554, 1044

\bibitem[\protect\citeauthoryear{Collins et al.}{2002}]{b08} Collins, J., A., Benjamin, R. A., Rand, R. J., 2002, ApJ, 578, 98
\bibitem[\protect\citeauthoryear{Cox et al.}{2005}]{b09} Cox, D. P., 2005, Annu. Rev. Astro. Astrophys., 43, 337
\bibitem[\protect\citeauthoryear{de Zeeuw et al.}{1999}]{b10} de Zeeuw, P. T., Hoogerwerf, R. de Bruijne J.H.J., 1999, ApJ,  117, 354

\bibitem[\protect\citeauthoryear{Draine \& Salpeter}{1979}]{b98} 
Draine, B.T., \& Salpeter, E.E. 1979, ApJ, 231, 438


\bibitem[\protect\citeauthoryear{Dwek E.}{1998}]{b98} Dwek, E., 1998, ApJ, 501, 643
\bibitem[\protect\citeauthoryear{Fraternali et al.}{2004}]{b11} Fraternali, F., Oosterloo, T.  Sancisi, R. 2004, A\&A, 424, 485
\bibitem[\protect\citeauthoryear{Fraternali \& Binney.}{2008}]{b49} Fraternali, F., Binney, J., 2008,  arXiv:0802.0496, accepted for publication in MNRAS
\bibitem[\protect\citeauthoryear{Grevesse \& Noels}{1993}]{grevesse93}
 Grevesse, N., \& Noels, A. 1993, in Orgin of the Elements, ed. N. Prantzos,
 E. Vangioni-Flam, \& M. Cass\'e, (Cambridge: Univ. Press), 15
\bibitem[\protect\citeauthoryear{Houck et al.}{1990}]{b19} Hernquist L.,1990, ApJ, 356, 359
\bibitem[\protect\citeauthoryear{Houck et al.}{1990}]{b12} Houck, J. C., \& Bregman, J. N. 1990, ApJ, 352, 506
\bibitem[\protect\citeauthoryear{Ibata et al.}{1994}]{b13} Ibata, R. A., Gilmore, G.,  Irwin, M. J. 1994, {\it Nature} 370, 194
\bibitem[\protect\citeauthoryear{Kompaneets et al.}{1960}]{b14} Kompaneets, A. S., 1960, Soviet Phys. Dokl., 5, 46


\bibitem[\protect\citeauthoryear{Jones et al.}{1996}]{b99} 
Jones, A.P., Tielens, G.G.M., \& Hollenbach, D.J. 1996, ApJ, 469, 740


\bibitem[\protect\citeauthoryear{Lu et al.}{1998}]{b16} Lu, L., Sargent, W. L. W., Savage, B. D., Wakker, B. P., Sembach, K. R., Oosterloo, T. A. 1998, AJ, 115, 162


\bibitem[\protect\citeauthoryear{Luck et al.}{2003}]{b260} 
  Luck R. E., Gieren W. P., Andrievsky S. M. et al. 
  2003, A\&A, 401, 939




\bibitem[\protect\citeauthoryear{Mac Low et al.}{1988}]{b19} Mac Low, M. M., McCray, R., 1988. ApJ, 324, 776
\bibitem[\protect\citeauthoryear{Matteucci et al.}{1986}]{b20} Matteucci, F., \& Greggio, L. 1986, A\&A, 154, 279
\bibitem[\protect\citeauthoryear{Matteucci}{2001}]{b20} Matteucci, F., 2001, The Chemical Evolution Of The Galaxy, Kluwer Academic Publishers.
\bibitem[\protect\citeauthoryear{Melioli et al.}{2001}]{b20} Melioli, 2008, A\&a Submitted .

\bibitem[\protect\citeauthoryear{Miyamoto et al.}{1975}]{b21} Miyamoto, M., \& Nagai, R., 1975, Publ. Astron. Soc. Japan, 27, 533


\bibitem[\protect\citeauthoryear{Murphy et al.}{2000}]{b22} Murphy, E. M., Sembach, K. R., Gibson, B. K., Shull, J. M., Savage, B. D., Roth, K. C., Moos, H. W., Green, J. C., York, D. G., Wakker, B. P., 2000, ApJ, 538, L35


\bibitem[\protect\citeauthoryear{Navarro et al.}{2001}]{b28} Navarro, J. D., Frenk, C. S., White S. D. M., 1996, ApJ, 462, 563

\bibitem[\protect\citeauthoryear{Oort et al.}{1970}]{b15} Oort, J. H. 1970, A\&A 7, 381


\bibitem[\protect\citeauthoryear{Richter et al.}{2001}]{b43}Richter, P., Sembach, K. R., Wakker, B. P., Savage, B. D., Tripp, T. M., Murphy, E. M., Kalberla, P. M. W., Jenkins, E. B., 2001, ApJ, 559, 318
\bibitem[\protect\citeauthoryear{Salpeter}{1955}]{b40} Salpeter, E. E., 1955, ApJ, 121, 161 
\bibitem[\protect\citeauthoryear{Sembach et al.}{1998}]{b40} Sembach, K. R., \& Oosterloo T. A. 1998, AJ, 115, 162 
\bibitem[\protect\citeauthoryear{Sembach et al.}{2001}]{b17} Sembach, K. R., Howk, J. C., Savage, B. D.,  Shull, J. M. 2001, AJ, 121, 992
\bibitem[\protect\citeauthoryear{Shapiro et al.}{1976}]{b18} Shapiro, P. R., \& Field, G. B. 1976, ApJ 205, 762





\bibitem[\protect\citeauthoryear{Tenorio-Tagle et al.}{1987}]{b23} Tenorio-Tagle, G., Bodenheimer, P., R{\'o}{\.z}yczka, M., 1987, A\&A, 182, 120
\bibitem[\protect\citeauthoryear{Tenorio-Tagle et al.}{1999}]{b50} Tenorio-Tagle, G., Silich, S. A., Kunth, D., Terlevich, E., Terlevich, R., 1999, Mon. Not. R. Astron. Soc., 309, 332

\bibitem[\protect\citeauthoryear{Wakker et al.}{2007}]{b24} Wakker, B. P., York, D. G., Howk, C., Barentine, J.C., Wilheim, R., Peletier, R. F., van Woerden, H., Beers, T. C., Ivezic, Z., Richter, P., Schwarz, U. J., 2007, ApJ, submitted
\bibitem[\protect\citeauthoryear{Wakker et al.}{1999}]{b24} Wakker, B. P., et al. 1999, Nature, 402, 388
\bibitem[\protect\citeauthoryear{ Woosley et al.}{1995}]{b25}  Woosley, S. E., Weaver, T. A., 1995, ApJ, 101, 181
\end{thebibliography}
\end{document}